# Ultrabroadband Coherent Perfect Absorption with Composite Graphene Metasurfaces

Wei Zou,[1] Tianjing Guo,[1,2,4*] Christos Argyropoulos[3,5*]

[1]*Department of Physics, School of Physics and Material Science, Nanchang University, Nanchang 330031, China*
[2]*Institute of Space Science and Technology, Nanchang University, Nanchang 330031, China*
[3]*Department of Electrical Engineering, The Pennsylvania State University, University Park, PA 16803, USA*
[4]*tianjing@ncu.edu.cn*
[5]*cfa5361@psu.edu*

**Abstract:** We investigate the design and performance of a new multilayer graphene metasurface for achieving ultrabroadband coherent perfect absorption (CPA) in the THz regime. The proposed structure comprises of three graphene patterned metasurfaces separated by thin dielectric spacer layers. The top and bottom metasurfaces have cross shape unit cells with varying sizes, while the middle graphene metasurface is square-shaped. This distinctive geometrical asymmetry and the presence of multiple layers within the structure facilitate the achievement of wideband asymmetric reflection under incoherent illumination. This interesting property serves as a crucial step towards achieving near-total absorption under coherent illumination across a broad frequency range. Numerical simulations demonstrate that the absorption efficiency surpasses 90% across an ultrabroadband frequency range from 2.8 to 5.7 THz, i.e., bandwidth of 2.9 THz. The CPA effect can be selectively tuned by manipulating the phase difference between the two incident coherent beams. Moreover, the absorption response can be dynamically adjusted by altering the Fermi level of graphene. The study also examines the influence of geometric parameters on the absorption characteristics. The results of this research work offer valuable insights into the design of broadband graphene metasurfaces for coherent absorption applications, and they contribute to the advancement of sophisticated optical devices operating in the THz frequency range.

## 1. Introduction

Coherent perfect absorption (CPA) is a fascinating phenomenon in the field of optics, where two counter-propagating beams illuminate on both sides an absorbing medium with appropriate phase and interfere in such a way resulting in the complete absorption of light [1-9]. This effect, also referred as time-reversed laser or anti-laser [2], has been the subject of extensive research since its initial demonstration in a Fabry-Perot cavity based on a silicon slab, where the trapped incident light underwent multiple reflections within the cavity until it was entirely absorbed due to destructive interference. The CPA devices provide additional tunability of absorption and interference by controlling the relative phase between the two incident coherent beams [10, 11]. While the concept of CPA has been explored in various contexts, including by using different nanostructures and subwavelength thin films [12-15], its practical application has been hindered by the limited operating frequency range, i.e., very narrowband response.

Recently, researchers have developed strategies to address this issue. One such strategy involves the simultaneous excitation of multiple resonances, enabling the superposition of several closely spaced absorption peaks. To achieve this, metamaterial unit cells with progressively varying geometric dimensions are integrated into a single plane, or multiple layers of metamaterial structures are vertically stacked, facilitating the simultaneous excitation of several resonances [16-19]. For instance, a dual-band polarization-independent CPA has been demonstrated based on a metal-graphene nanostructure, which features golden nanorings of varying sizes dispersed on a monolayer of graphene. Additionally, a design incorporating four sizes of columnar metal patches has been reported to operate within the mid-infrared range, capable of achieving multiband and broadband CPA by optimizing the radius of the metal patches and the thickness of the dielectric layer [19]. Through meticulous optimization of the structural parameters, these resonances can be closely spaced, enhancing the absorption bandwidth of the system. Another approach to broadening the CPA bandwidth is by significantly reducing the size of the absorbing medium [19, 20]. Studies have shown that CPA response with a rather large bandwidth can be realized using an ultrathin, heavily doped silicon slab [14]. By combining the aforementioned two strategies, a general method that utilizes epsilon-near-zero (ENZ) multilayer films was demonstrated to achieve CPA, where each layer is characterized by its ultrathin thickness [21]. However, despite these advancements, the bandwidth of the reported CPA responses remains limited for practical applications.

In this study, a multilayer graphene composite structure is proposed to achieve ultrabroadband CPA. The proposed structure consists of three layers made of patterned graphene, each separated by ultrathin dielectric spacer layers. The top and bottom graphene layers consist of cross-shaped unit cell structures, whereas the middle layer is square-shaped. Notably, the entire thickness of the composite structure is just 2.4 um, much smaller than the operating wavelength. We first derive the CPA condition using the scattering matrix and then perform numerical simulations of the proposed structure under incoherent illumination. The simulation results show that an absorption of nearly 50% is achievable across a broad frequency spectrum for both illumination directions, suggesting the potential for near-total absorption when employing two coherent beams. Then we conduct simulations under coherent counterpropagating illuminations, revealing near-total absorption within the range of 3.1 to 5.2 THz, thereby demonstrating the ultrabroadband CPA effect. Moreover, this CPA effect can be dynamically modulated by adjusting the phase difference between the two incident coherent beams. We further elucidate the underlying physical mechanisms of the presented CPA effect through a comprehensive analysis of the induced magnetic field and surface current distributions. In addition, we investigate the impact of key geometric parameters on the absorption response and examined the role of the graphene's Fermi level in modulating the absorption characteristics. These findings provide critical insights into the tunability of the CPA response. Our research paves the way for the future design and optimization of broadband tunable graphene metasurfaces, specifically for applications involving coherent absorption in the THz regime.

## 2. Theory and Design

A graphene monolayer can be considered as an effectively ultrathin conductive surface, with its conductivity being linked to the Fermi energy EF that can be altered by chemical or electrical doping. This material parameter can be modeled using the well-known Kubo formula [22], which consists of two terms: the intraband contribution σintra and the interband contribution σinter. In the regime where the frequency $\omega$ is below a certain threshold $\hbar\omega = 2E_F\hbar$, the interband term σinter is negligible, while the intraband term σintra dominates [23]. In the THz regime, the conductivity of graphene is primarily determined by the intraband contribution, which can be evaluated using a Drude-like expression [24]:

$$\sigma = \sigma_{intra} = \frac{\sigma_0}{1+j\omega\tau}, \tag{1}$$

where

$$\sigma_0 = \frac{e^2 \tau k_B T}{\pi \hbar^2}[\frac{E_F}{k_B T} + 2ln(\exp\{-\frac{E_F}{k_B T}\}+1)]. \tag{2}$$

In Eq. (2) $\sigma_0$ is the DC conductivity, $e$ is the electron charge, $k_B$ is the Boltzmann constant, $T$ is the temperature, and $\tau$ is the electron relaxation time. Throughout this work, we adopt Eq. (1) as the surface conductivity model for the monolayer graphene sheet, since the proposed structure is investigated at the THz regime.

Figure 1(a) presents a schematic illustration of the coherent absorption set-up and the proposed composite metasurface structure. This figure depicts the scenario where two coherent light beams traverse the multilayer metasurface from opposite directions. The interplay between these two beams, mediated by their relative phase difference, governs the occurrence and efficiency of coherent absorption. The unit cell of the broadband CPA composite metasurface design, shown in Fig. 1(b), consists of three layers of patterned graphene metasurfaces separated by thin dielectric spacer layers. The uppermost and lowermost graphene metasurfaces consist of varying size cross patterns, whereas the middle layer is square-shaped. In the context of coherent illumination, there are two counter-propagating coherent beams $E_{in1}$ and $E_{in2}$, that impinge vertically upon an absorptive structure, as depicted in Fig. 1(b). The intensities of the output beams, or the scattered beams, from each side are denoted as $E_{out1}$ and $E_{out2}$, respectively. The relationship between the incident and output beams can be mathematically represented as [25]:

$$\begin{bmatrix} E_{out1} \\ E_{out2} \end{bmatrix} = S \begin{bmatrix} E_{in1} \\ E_{in2} \end{bmatrix} = \begin{bmatrix} r_1 & t_1 \\ t_2 & r_2 \end{bmatrix}\begin{bmatrix} E_{in1} \\ E_{in2} \end{bmatrix}. \tag{3}$$

Here, $S$ is the scattering matrix which is derived by simulations where the Maxwell's equations are numerically solved. The parameters $r_1$, $t_1$, $r_2$ and $t_2$ are the reflection and transmission coefficients in the case of a single beam of $E_{in1}$ and $E_{in2}$. It is important to note that, under this configuration, the transmission coefficient is identical for both incident directions (i.e., $t = t_1 = t_2$), given that reciprocity remains unbroken in this setup. The asymmetry in reflection arises from the different size of the cross-shaped graphene patches arranged on the opposite sides of the designed structure.

To quantitatively investigate CPA, the coherent absorption metric is defined as [25, 26]:

$$A_{co} = 1 - \frac{P_{out1} + P_{out2}}{P_{in1} + P_{in2}} = 1 - \frac{|E_{out1}|^2 + |E_{out2}|^2}{|E_{in1}|^2 + |E_{in2}|^2}. \tag{4}$$

As a result, the CPA performance corresponds to zero output intensities: $|E_{out1}|^2 + |E_{out2}|^2 = 0$ where $A_{co}=1$. To simplify the derivation of the CPA condition, we assume symmetric in terms of amplitude illuminations $E_{in1} = E_{in2}e^{i\phi}$, which are used throughout this work, where $\phi$ represents the phase difference between the two incident waves. From Eq. (4) with $A_{co}=1$, the CPA condition can be obtained:

$$\begin{cases} r_1 = r_2 e^{i2\phi} \\ t = -r_2 e^{i\phi} \end{cases}. \tag{5}$$

By properly designing the system, we can achieve certain $r_1$, $r_2$ and t to satisfy the CPA condition given by Eq. (5). As Eq. (5) also shows, by employing asymmetrical configurations, it is possible to achieve CPA under conditions of symmetric illumination. Finally, it should be noted that we always convert the calculated coherent absorption metric $A_{co}$ to decibel (dB) units to facilitate a comprehensive and meaningful comparison for different illumination phase differences.

The design of the proposed CPA device is refined through simulations conducted by COMSOL Multiphysics [27], a proprietary electromagnetic simulation software that employs the finite element method (FEM). Given the diverse shapes of the graphene patches on each layer, a three-dimensional (3D) simulation domain is essential, instead of a two-dimensional (2D) model. The optical conductivity of graphene, described by Eq. (1), is utilized in these simulations. To replicate the periodic nature of the graphene patches, periodic boundary conditions (PBCs) are applied in the x and y directions, as shown in Fig. 1(b). In the z direction, port boundaries are established to realize the incoming plane waves, including the downward and upward illuminations. This setup ensures that the simulations accurately reflect the device's intended functionality and optimize its performance. The simulation details provided above are specific to the scenario where a single light beam is being modeled. For coherent illumination, the built-in solvers within COMSOL Multiphysics compute by solving Maxwell's equations the electric and magnetic fields across the entire simulation domain, allowing us to analyze the fields and power by placing probes at the output ports. We can quantitatively calculate the coherent absorption by using the formula in Eq. (4), where the output power is determined by summing the detected power and dividing over the input power. Here, this summation is necessary due to the different direction between the input and output energy vectors.

Considering a potential experimental implementation of the proposed CPA device, an overview of the manufacturing process is provided [28, 29]: initially, uniform graphene sheets are grown on pre-positioned Silicon Carbide (SiC) / Silicon (Si) pseudo-substrates via chemical vapor deposition (CVD); subsequently, these graphene sheets are transferred to a silicon dioxide wafer substrate using polymethyl methacrylate assistive techniques [30]. The patterning of the graphene sheets can be achieved through electron-beam lithography (EBL) [31, 32] or direct CVD synthesis [29, 33]; for instance, in [33] tightly packed graphene nanodisks with a diameter of 60 nm and a spacing of 30 nm was successfully fabricated, while in [29] graphene gratings with a width of at least 100 nanometers were produced. Top-down methods, such as etching or cutting, are physical processes that can produce wafer-scale graphene compatible with large-scale integration [34, 35]. Considering that the dimensions of the patterned structures in this study are within the micrometer scale, the fabrication of such graphene metasurfaces is deemed practical. Thereafter, the same treatment is repeated on the opposite side of the silicon wafer to obtain the patterned graphene metasurface on the reverse side. An ionic gel [31, 36-38] with a relative permittivity of 1.82 is used as the dielectric material to realize the graphene metasurface gating configuration, and two gold contacts are fabricated on top to apply voltage, dope the graphene, and tune the absorption performance, as depicted in Fig. 1(a). Finally, a layer of the same silicon dioxide wafer is covered, followed by the transfer and re-patterning of graphene, the covering of the ionic gel layer, and the construction of gold contacts, thus completing the fabrication of the desired multilayer composite graphene metasurface. Note that since the ionic gel layer is extremely thin and has a low dielectric constant, its impact on the system performance is minimal and is thus ignored in simulations. Creating a realistic multilayer metasurface structure might be challenging, but it is feasible with the guidance provided above.

Moreover, the measurement of absorption is equally crucial as the fabrication process. For effective achievement of the CPA effect, it is necessary to have a variable path delay and an attenuator, as it has been experimentally demonstrated in [2]. These components are employed to regulate the phase difference between the two incoming beams and to ensure that the intensities of the beams are equal, respectively. On each side of the structure, two beam splitters are utilized to simultaneously introduce input beams and export output beams. The two output beams are then combined by another beam splitter into a high-resolution spectrometer. The intensity of each output beam can be determined by simply blocking the other beam. To ascertain the total output intensity, both output beams are left unobstructed. Interference is eliminated by positioning all the beam splitters such that the paths of the output beams into the spectrometer differ by more than the coherence length of the laser. We can construct a similar experimental setup in the THz range, utilizing continuous wave tunable frequency THz sources, as demonstrated in [39]. In the proposed structure, graphene Fermi level can be changed by tuning the voltage applied to the gate, thereby achieving tunable absorption. In the case of moderately doped graphene, i.e., $E_F \gg KBT$, the Fermi energy can be obtained by the formula $E_F = \hbar v_F \sqrt{\pi C V_g / e}$ [40], where $\hbar$ is the Planck constant, $v_F$ is the Fermi velocity and $V_g$ is the gate voltage between the electrodes. The electrostatic capacitance per unit area can be described as $C = \varepsilon_r \varepsilon_0 / d$, where $\varepsilon_r$ and $d$ is the static permittivity and thickness of dielectric layer, respectively.

In this study, our aim is to achieve an ultra-broadband CPA response by designing a composite graphene structure with high structural asymmetry. To further enhance this asymmetry, we utilize two cross-shaped graphene patches of varying sizes on different sides of the structure. The multilayer metasurface structure is optimized through single illumination simulations. To meet the CPA condition expressed by Eq. (5) in a broad frequency range, the period $p$ of the unit cell is set equal to 8 um, with a distance $d$ of 1.2 um between adjacent graphene metasurfaces, and a dielectric layer with relative permittivity $\varepsilon_r$ of 2.25, akin to that of silica. Note that the CPA performance could be further improved by incorporating lossy dielectric materials into the proposed design. The dimensions of the bottom large cross-shaped graphene patches have length $L_1$ of 7.2 um and width $W_1$ of 2.8 um. The middle square graphene patches have a side length $W_2$ of 2.8 um, while the uppermost small cross-shaped graphene patches features a length $L_3$ of 4.6 um and width $W_3$ of 2.4 um. Throughout this work, the Fermi level of the graphene monolayer is consistently set at 0.5 eV, unless explicitly specified otherwise.

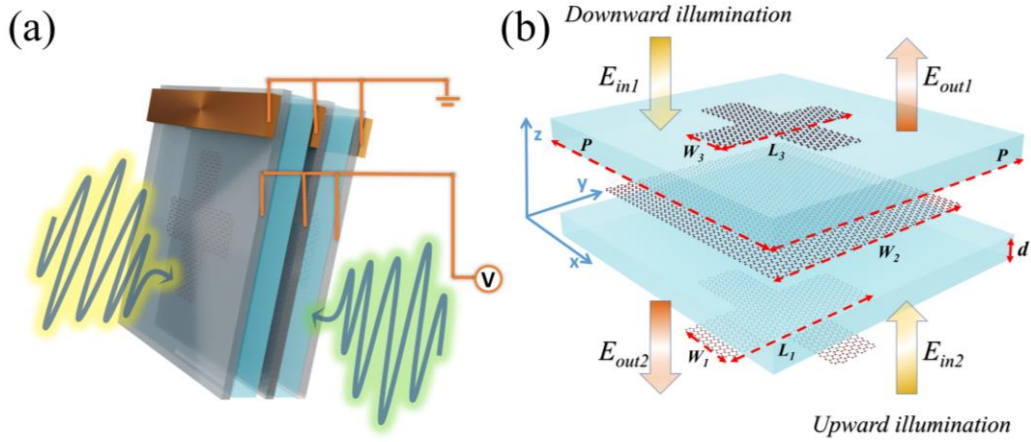

Figure. 1. (a) Schematic diagram of coherent absorption achieved with the proposed multilayer composite metasurface structure. Two coherent optical beams strike the structure from opposite directions at normal angle of incidence. (b) Unit cell of the proposed composite metasurface structure and its geometrical parameters.

## 3. Results and Discussion

### 3.1 Incoherent and coherent absorption

We begin by calculating the spectra of transmission, reflection, and absorption under incoherent illumination from two different directions. When a single beam of light is incident on the structure at normal incidence, some of the energy is absorbed, while the rest is reflected or transmitted. The results for downward and upward single beam illuminations are presented in Figs. 2(a) and (b), respectively. As explained in Section 2, transmission is the same for both illumination directions. The reflection coefficient, across a wide frequency range (2.8-5.6 THz), is similar for both illuminations, as shown in Fig. 2. According to Eq. (5), the reflection coefficient magnitudes $|r_1|$ and $|r_2|$ from opposite sides should be the same to satisfy the CPA condition. Importantly, the absorption from the two illumination directions is also close to 50% within this broad frequency range, as shown by blue lines in Fig. 2(a) and (b). These results provide a foundation for realizing broadband CPA response with the proposed structure. The 50% absorption is the maximum value that can be achieved by an ultrathin symmetric subwavelength structure [41]. However, at certain frequency points, such as 3.1 THz under downward illumination and 5.2 THz under upward illumination, the absorption slightly exceeds the 50% limit due to the critically asymmetric design of the proposed structure, which utilizes the coupling effect between different graphene layers [42].

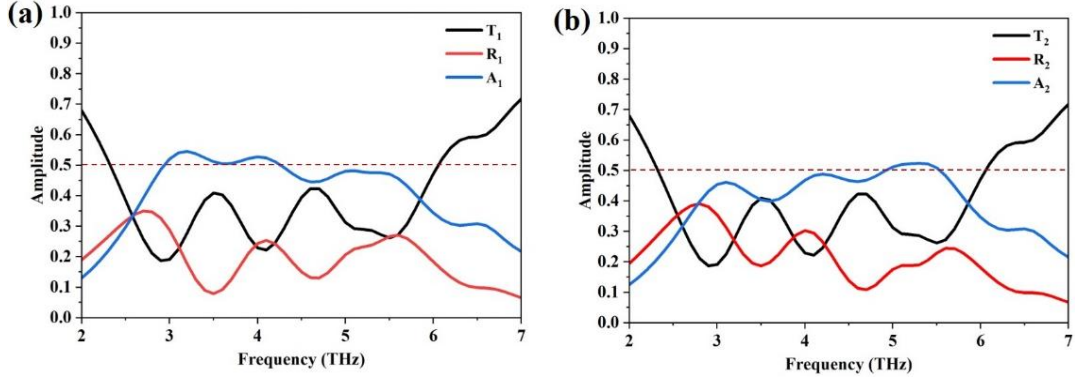

Figure. 2. Computed transmission, reflection and absorption spectra of the proposed multilayer composite metasurface structure under single illumination from opposite directions: (a) downward and (b) upward illumination.

When two coherent beams of equal intensity illuminate the structure from opposite directions, they undergo interference, resulting in coherent absorption. The nature of this absorption is contingent upon the relative phase difference between the two counter-propagating beams. In the calculation of Fig. 3, we illustrate the relationship between the coherent absorption (computed by Eq. (4)) and phase difference. It is evident that, when the phase difference is precisely 0°, the total incident energy is perfectly absorbed, indicating that the critical point for CPA is achieved. Under this condition, the interference between the two beams reduces their scattering, thereby enhancing the absorption process. The system can be detuned from the CPA condition by adjusting the phase difference. As the phase difference is incrementally increased, the coherent absorption progressively diminishes, eventually reaching zero at a phase difference of 180°. It is particularly noteworthy that, even with a phase difference of 50°, the coherent absorption can still attain a substantial level of 80%.

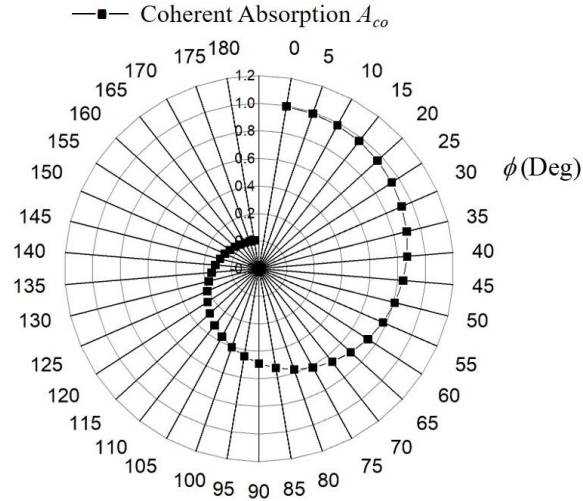

Figure 3. Computed coherent absorption metric $A_{co}$ of the proposed multilayer structure by varying the phase difference between the two incident counterpropagating waves. The operating frequency is 3.1 THz.

We conduct a detailed analysis of coherent absorption by systematically adjusting the operating frequency and the phase difference between the two incident waves, as depicted in Fig. 4(a). By keeping the phase of the incident wave $E_{in1}$ at 0°, we sweep the phase of the incident wave $E_{in2}$ from 0° to 180°.

Subsequently, we fix the phase of the incident wave $E_{in2}$ at 0°, while we change the phase of the incident wave $E_{in1}$ from 0° to 180°. As a result, the phase difference varies from -180 to 180° while the frequency always changes from 1 to 9 THz. Our findings indicate that the sign of the phase difference does not affect the coherent absorption performance. Coherent absorption efficiencies of ≥90% are achieved when the phase difference falls within the range of -30° to 30°. Moreover, an ultrabroadband near total absorption spectrum is achieved from 3.1 to 5.2 THz, with a bandwidth of 2.1 THz, which exceeds the performance of other relevant designs [43, 44], considering the ultrathin thickness of the proposed structure. Importantly, coherent absorption efficiencies of ≥90% are maintained across this broad bandwidth, demonstrating the robustness and versatility of the absorption mechanism. For a more visually intuitive understanding, Fig. 4(b) displays the coherent absorption spectra for phase differences of 0° and 60°. These spectra further corroborate the robustness of the proposed multilayer graphene structure against fluctuations in phase. This characteristic is particularly advantageous for applications in optical device design and sensor technology, where precise manipulation of the incident light phase is essential for optimal performance. Moreover, the graph in Fig. 4(b) highlights the broad bandwidth of coherent absorption. The structure exhibits a remarkable ability to maintain high coherent absorption efficiency across a wide range of phase differences and operating frequencies. This dual robustness against phase variations and frequency shifts further emphasizes the potential utility of the structure in the aforementioned fields. The proposed multilayer graphene structure demonstrates exceptional resilience in maintaining high coherent absorption efficiency, making it a promising candidate for various applications in optical device designs.

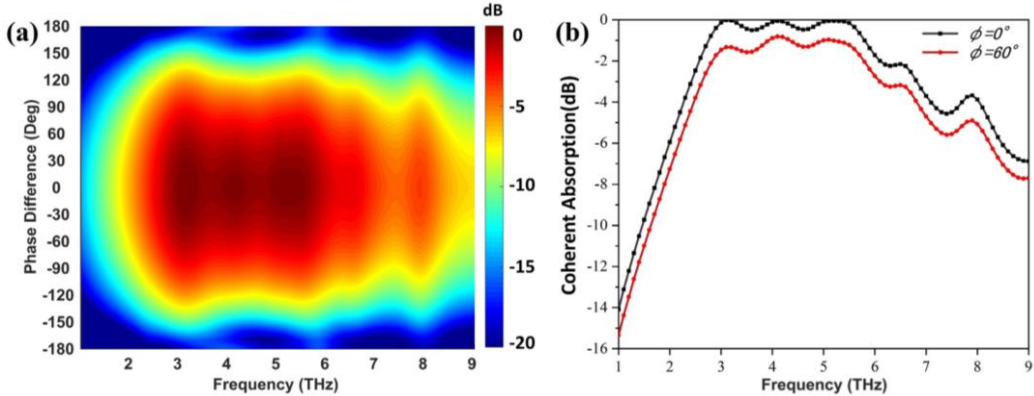

Figure 4. (a) Computed coherent absorption metric *Aco* as a function of the operating frequency and the phase difference between two coherent beams. (b) Coherent absorption metric *Aco* as a function of the operating frequency when the phase difference is set as 0° and 60°.

In order to provide physical insights into the electromagnetic response of the proposed structure, we analyze the magnetic field distribution induced at the proposed CPA metasurface structure under incoherent illumination. Figures 5(a) and (b) show the magnetic field $H_z$ and surface current distribution of each graphene layer in the proposed structure at $f_1$ = 3.1 THz and $f_2$ = 5.2 THz, respectively, when the structure is excited with downward illumination. The magnetic field $H_z$ is mainly concentrated at the intersections and edges of the cross-shaped graphene metasurfaces for both operating frequencies. A notable observation in Figs. 5(a) and 5(b) is the occurrence of a magnetic quadrupole resonance at both $f_1$ and $f_2$ frequencies [45-47]. Interestingly, the magnetic field distribution of the large cross graphene metasurface is opposite to that of the square and the small cross graphene metasurface at the top, at the $f_1$ frequency, as depicted in Fig. 5(a). This discrepancy may be attributed to the fact that the topmost graphene metasurface is the first to be excited by electromagnetic radiation. The lower patterned graphene layers, however, are influenced by both the incident radiation and the resonance radiation from the structure, resulting in the opposite magnetic field distribution. At the $f_2$ frequency shown in Fig. 5(b), the magnetic field distribution in the lower two layers is predominantly concentrated at the edges of the graphene patches, forming a consistent $H_z$ distribution. This observation provides valuable insights into the behavior of the graphene layers within the proposed metasurface structure, contributing to a better understanding of the structure's functionality. By the precise analysis of the magnetic field and surface current distribution within the proposed CPA metasurface structure under incoherent illumination, we verify that each layer can be designed to interact with the incoming waves in a way that complementary absorption is achieved across a broad frequency range. The detailed investigation of the magnetic field and surface current distribution has shed light on the intricate interaction between the structure's design and its ability to manipulate electromagnetic waves, ultimately enhancing the coherent absorption bandwidth.

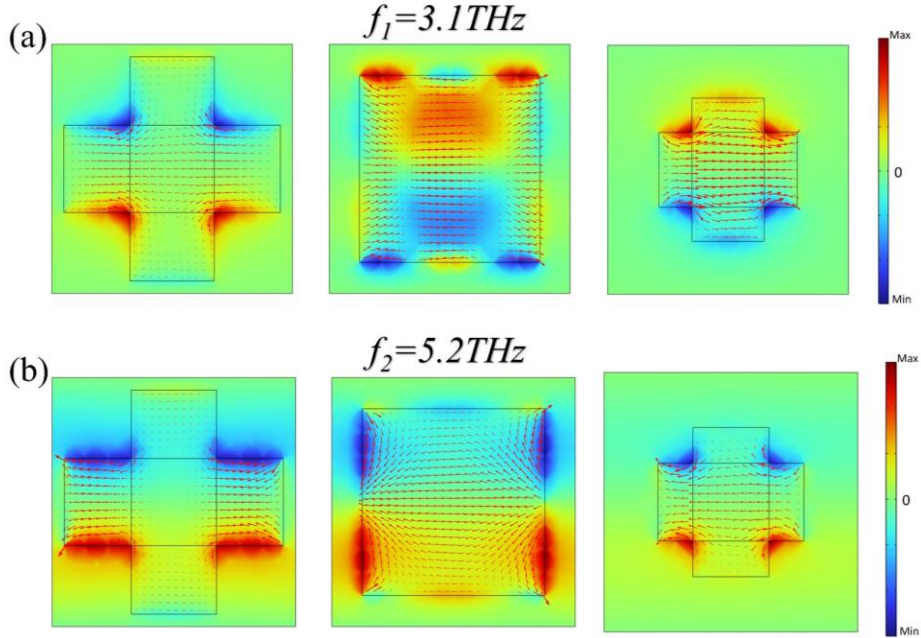

Figure 5. Computed distribution of the magnetic field Hz and surface current across three graphene metasurface unit cells under downward illumination at frequencies (a) $f_1$ = 3.1 THz and (b) $f_2$ = 5.2 THz. In these simulations, the small cross graphene patch is excited first.

*3.2 Tunable coherent absorption*

We investigate the impact of various geometrical parameters on the absorption spectra of the designed CPA composite structure. Figure 6 displays the computed results, where $L_1$ and $L_3$, respectively, represent the lengths of the bottom and uppermost cross-shaped graphene patches. These dimensions are varied independently while all other structural parameters are kept constant. The phase difference between the two coherent beams is set to 0°. It is evident that broadband coherent absorption can still be achieved with different structural parameters. As depicted in Fig. 6(a), the bandwidth of the coherent absorption with values exceeding 90% gradually increases as $L_1$ increases, reaching a maximum at 7.3 μm. Subsequently, the absorption spectrum exhibits oscillatory behavior when $L_1$ further increases, leading to the formation of multiple absorption peaks. By modifying the length $L_3$, we compute the coherent absorption as a function of $L_3$ and the operating frequency, as shown in Fig. 6(b). This result demonstrates that the influence of $L_3$ on the bandwidth of ≥90% coherent absorption values is minimal, as changes in $L_3$ scarcely affect the magnetic quadrupole resonance triggered by the structure when other structural parameters are unchanged. It is evident that the adjustment of $L_3$ has a limited impact on the CPA performance, while the transition from broadband absorption to multi-peaked absorption can be achieved through the manipulation of the parameter $L_1$.

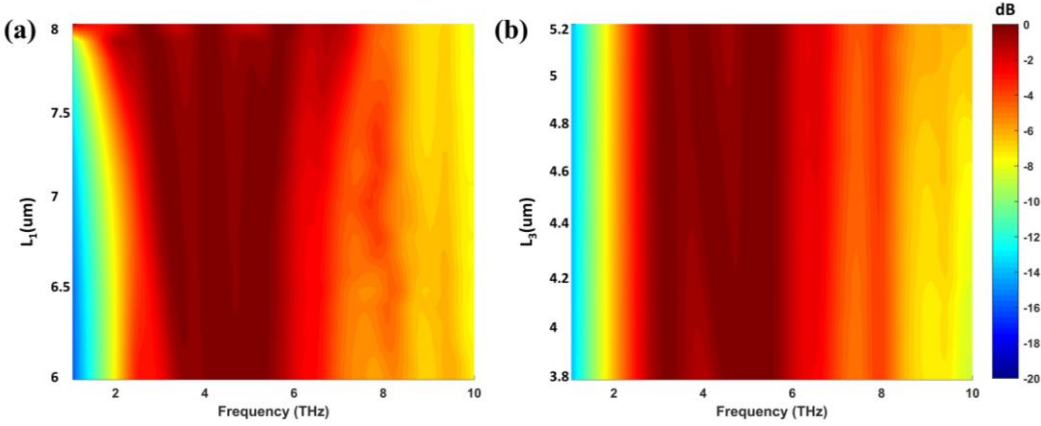

Figure 6. Computed coherent absorption metric $A_{co}$ as function of the operating frequency and structure parameters: (a) the length of the bottom large cross-shaped graphene patch, and (b) the length of the uppermost small cross-shaped graphene patch.

Tunability is a distinctive attribute of graphene, enabling the dynamic control of the coherent absorption in the proposed device by manipulating its Fermi level. To examine the impact of the Fermi level on the CPA performance, we plot the distribution of the coherent absorption coefficient as the Fermi level $E_F$ varies from 0 to 1 eV, and the incident frequency ranges from 1 to 10 THz. These results are presented in Fig. 7. Initially, we observe a notable blue shift in the resonant frequency with the elevation of the Fermi level. This phenomenon aligns with the established correlation between the resonant frequency and the Fermi level in graphene patches, which can be concisely represented as $f_r \propto \sqrt{E_F/L}$ [48, 49], with $L$ denoting the side length of the graphene patch or the structure geometry. Furthermore, from Fig. 7, we deduce that an ultra-wideband CPA with an absorption coefficient exceeding 90% can be realized when the Fermi level exceeds 0.2 eV, which corresponds to a moderate doping level. As a result, for the design of the current adjustable and broadband CPA device, it is unnecessary to employ graphene with excessively high doping values. This implies that there is no requirement to substantially raise the applied gate voltage proportionally to the Fermi level value to achieve the desired performance of the broadband CPA. Note that the necessary voltage to achieve Fermi level equal to 0.2 eV is approximately 7.2 V, since the thickness of the dielectric layer in our structure is $d = 1.2$ um, and the relative permittivity of the used dielectric layer is $\varepsilon_r = 2.25$. This voltage value is relatively low and will not result in a dielectric breakdown. Importantly, the system's detuning from the CPA condition can be achieved by altering the Fermi level, for instance, through the electrochemical potential of a metallic gate serving as the back-reflector. Additionally, with the increase of graphene's Fermi level, we note that the bandwidth of the proposed structure widens progressively. It is possible to finely tune the bandwidth and resonance frequency of the broadband CPA by precisely adjusting the Fermi level of graphene.

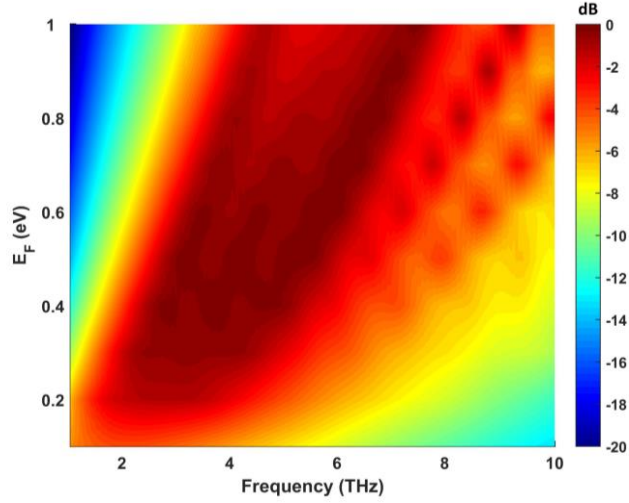

Figure 7. Computed coherent absorption metric *Aco* as a function of operating frequency and Fermi level of graphene. A significant blue shift is observed in the resonant frequency with the increase of graphene's Fermi level.

It is crucial to assess the range of incident angles over which the proposed structure can achieve near-perfect absorption, particularly in the context of nonplanar illumination scenarios [50]. We consider downward illumination as normal incidence, hence maintaining a constant incident angle at $\theta_1 = 0°$. We then vary the incident angle $\theta_2$ from 0° to 30° for upward illumination and calculate the coherent absorption coefficient as a function of operating frequency. As depicted in Fig. 8(a), the phenomenon of near total absorption persists over a substantial range of incidence angles, up to 20°. At the incident angle of 30°, we observe a slight blue shift in the resonance frequency compared to that at smaller incident angles $\theta_2$. Similarly, we maintain normal upward illumination and investigate the impact of incident angle $\theta_1$ on the absorption performance. The results are presented in Fig. 8(b). It is intriguing to observe that the incident angle $\theta_1$ exhibits a limited influence on the coherent absorption of the proposed structure. This can be attributed to the different size of the graphene patches on opposite sides of the structure; the smaller graphene patch exerts a lesser effect on downward illumination with an incident angle $\theta_1$. The results in Fig. 8 illustrate that the coherent absorption of the proposed multilayer structure is robust against variations in both incident angles, $\theta_1$ and $\theta_2$, demonstrating its versatility under different oblique incident light conditions.

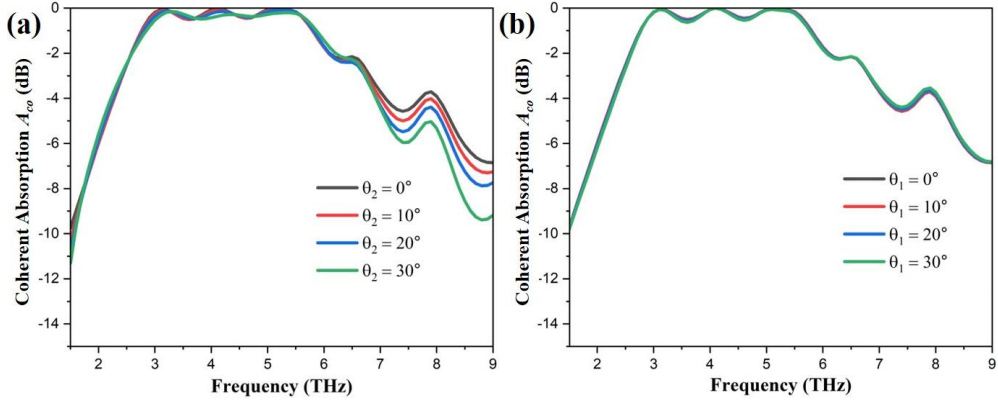

Figure 8. Computed incident-angle dependence of the coherent absorption metric $A_{co}$ of the proposed metasurface structure. (a) The coherent absorption metric under various incident angles $\theta_2$ while keeping the incident angle $\theta_1$ constant at 0°. (b) The coherent absorption metric under various incident angles $\theta_1$ while keeping the incident angle $\theta_2$ constant at 0°.

## 4. Conclusions

In summary, we have demonstrated the realization of an ultrabroadband CPA performance in the THz frequency range using a meticulously designed multilayer graphene composite metasurface. This structure comprises three distinct layers of patterned graphene metasurfaces, each separated by ultrathin dielectric spacers, and is characterized by a high degree of asymmetry in its geometric configuration. The strategic arrangement of these layers and their asymmetric geometries are pivotal in achieving near-total absorption across an extensive frequency band. Our analysis of the absorption characteristics under both incoherent and coherent illuminations reveals the potential for achieving near-total coherent absorption over a broad frequency spectrum. Notably, the absorption efficiency exceeds 90% across an ultrabroad frequency range from 2.8 to 5.7 THz, with a bandwidth of 2.9 THz. More impressively, near-total (almost 100%) coherent absorption is demonstrated from 3.1 to 5.2 THz, showcasing an ultrabroadband CPA effect. This CPA effect can be dynamically tuned by adjusting the phase difference between the two coherent incident beams. To elucidate the underlying mechanisms of this broadband CPA effect, we conducted a detailed analysis of the magnetic field and surface current distributions. We also explored the influence of key geometric parameters, specifically the lengths of the two cross-shaped graphene patches, on the CPA performance. Furthermore, we investigated the role of the graphene's Fermi level in modulating the absorption response, providing insights into the tunability of the CPA response. These comprehensive findings not only underscore the effectiveness of our design in achieving ultrabroadband CPA performance but also highlight the potential applications of such graphene composite metasurfaces in advanced technologies such as compact all-optical switches and sensitive coherent light detectors. Our work offers valuable guidance for the future design and optimization of broadband tunable graphene metasurfaces tailored for coherent absorption applications in the THz regime.

**Funding.** This work was supported by the National Science Foundation of China (NSFC) under Grant No. 12104203, the National Science Foundation under Grant No. DMR-2224456, and Jiangxi Double-Thousand Plan (Grant No. jxsq2023101069).

**References**

1. Y. D. Chong, L. Ge, H. Cao, and A. D. Stone, "Coherent Perfect Absorbers: Time-Reversed Lasers," Physical review letters **105** (2010).
2. W. Wan, Y. Chong, L. Ge, H. Noh, A. D. Stone, and H. Cao, "Time-Reversed Lasing and Interferometric Control of Absorption," Science **331**, 889-892 (2011).
3. M. Kang, Y. D. Chong, H.-T. Wang, W. Zhu, and M. Premaratne, "Critical route for coherent perfect absorption in a Fano resonance plasmonic system," Applied Physics Letters **105** (2014).
4. L. Liu, W. Liu, and Z. Song, "Ultra-broadband terahertz absorber based on a multilayer graphene metamaterial," Journal of Applied Physics **128**, 093104 (2020).
5. S. Quader, M. R. Akram, F. Xiao, and W. Zhu, "Graphene based ultra-broadband terahertz metamaterial absorber with dual-band tunability," Journal of Optics **22** (2020).
6. H. Hajian, A. Ghobadi, B. Butun, and E. Ozbay, "Active metamaterial nearly perfect light absorbers: a review Invited," Journal of the Optical Society of America B-Optical Physics **36**, F131-F143 (2019).
7. S. Wu, and J.-S. Li, "Hollow-petal graphene metasurface for broadband tunable THz absorption," Appl. Opt. **58**, 3023-3028 (2019).
8. S. Wu, D. Zha, L. Miao, Y. He, and J. Jiang, "Graphene-based single-layer elliptical pattern metamaterial absorber for adjustable broadband absorption in terahertz range," Physica Scripta **94** (2019).
9. E. B. Bosdurmaz, H. Hajian, V. Ercaglar, and E. Ozbay, "Graphene-based metasurface absorber for the active and broadband manipulation of terahertz radiation," Journal of the Optical Society of America B-Optical Physics **38**, C160-C167 (2021).
10. Z. Zhang, M. Kang, X. Zhang, X. Feng, Y. Xu, X. Chen, H. Zhang, Q. Xu, Z. Tian, W. Zhang, A. Krasnok, J. Han, and A. Alu, "Coherent Perfect Diffraction in Metagratings," Advanced Materials **32** (2020).
11. M. A. Kats, D. Sharma, J. Lin, P. Genevet, R. Blanchard, Z. Yang, M. M. Qazilbash, D. N. Basov, S. Ramanathan, and F. Capasso, "Ultra-thin perfect absorber employing a tunable phase change material," Applied Physics Letters **101** (2012).
12. Y. Fan, Z. Liu, F. Zhang, Q. Zhao, Z. Wei, Q. Fu, J. Li, C. Gu, and H. Li, "Tunable mid-infrared coherent perfect absorption in a graphene meta-surface," Scientific Reports **5** (2015).
13. S. Dutta-Gupta, O. J. F. Martin, S. D. Gupta, and G. S. Agarwal, "Controllable coherent perfect absorption in a composite film," Optics Express **20**, 1330-1336 (2012).
14. M. Pu, Q. Feng, M. Wang, C. Hu, C. Huang, X. Ma, Z. Zhao, C. Wang, and X. Luo, "Ultrathin broadband nearly perfect absorber with symmetrical coherent illumination," Optics Express **20**, 2246-2254 (2012).
15. J. Zhang, K. F. MacDonald, and N. I. Zheludev, "Controlling light-with-light without nonlinearity," Light-Science & Applications **1** (2012).

16. T. Zhou, S. Wang, Y. Meng, S. Wang, Y. Ou, H. Li, H. Liu, X. Zhai, S. Xia, and L. Wang, "Dynamically tunable multi-band coherent perfect absorption based on InSb metasurfaces," Journal of Physics D-Applied Physics **54** (2021).
17. W. Kang, Q. Gao, L. Dai, Y. Zhang, H. Zhang, and Y. Zhang, "Dual-controlled tunable terahertz coherent perfect absorption using Dirac semimetal and vanadium dioxide," Results in Physics **19** (2020).
18. J. Si, Z. Dong, X. Yu, and X. Deng, "Tunable polarization-independent dual-band coherent perfect absorber based on metal-graphene nanoring structure," Optics Express **26**, 21768-21777 (2018).
19. S. Huang, Z. Xie, W. Chen, J. Lei, F. Wang, K. Liu, and L. Li, "Metasurface with multi-sized structure for multi-band coherent perfect absorption," Optics Express **26**, 7066-7078 (2018).
20. T. Guo, and C. Argyropoulos, "Tunable and broadband coherent perfect absorption by ultrathin black phosphorus metasurfaces," Journal of the Optical Society of America B-Optical Physics **36**, 2962-2971 (2019).
21. T. Y. Kim, M. A. Badsha, J. Yoon, S. Y. Lee, Y. C. Jun, and C. K. Hwangbo, "General Strategy for Broadband Coherent Perfect Absorption and Multi-wavelength All-optical Switching Based on Epsilon-Near-Zero Multilayer Films," Scientific Reports **6** (2016).
22. V. P. Gusynin, S. G. Sharapov, and J. P. Carbotte, "Magneto-optical conductivity in graphene," Journal of Physics-Condensed Matter **19** (2007).
23. G. W. Hanson, "Dyadic Green's functions and guided surface waves for a surface conductivity model of graphene," Journal of Applied Physics **103** (2008).
24. L. A. Falkovsky, "Optical properties of graphene," in *International Conference on Theoretical Physics (Dubna-Nano2008)*(JINR, Bogoliubov Lab Theoret Phys, Dubna, RUSSIA, 2008).
25. Z. J. Wong, Y.-L. Xu, J. Kim, K. O'Brien, Y. Wang, L. Feng, and X. Zhang, "Lasing and anti-lasing in a single cavity," Nature Photonics **10**, 796-801 (2016).
26. S. Longhi, "*PT*-symmetric laser absorber," Physical Review A **82** (2010).
27. Comsol Multiphysics, http://www.comsol.com/.
28. X. Yan, T. Li, G. Ma, J. Gao, T. Wang, H. Yao, M. Yang, L. Liang, J. Li, J. Li, D. Wei, M. Wang, Y. Ye, X. Song, H. Zhang, C. Ma, Y. Ren, X. Ren, and J. Yao, "Ultra-sensitive Dirac-point-based biosensing on terahertz metasurfaces comprising patterned graphene and perovskites," Photonics Research **10**, 280-288 (2022).
29. D. A. Katzmarek, A. Mancini, S. A. Maier, and F. Iacopi, "Direct synthesis of nanopatterned epitaxial graphene on silicon carbide," Nanotechnology **34** (2023).
30. H. Hu, X. Yang, F. Zhai, D. Hu, R. Liu, K. Liu, Z. Sun, and Q. Dai, "Far-field nanoscale infrared spectroscopy of vibrational fingerprints of molecules with graphene plasmons," Nature Communications **7** (2016).
31. Z. Fang, Y. Wang, A. E. Schather, Z. Liu, P. M. Ajayan, F. Javier Garcia de Abajo, P. Nordlander, X. Zhu, and N. J. Halas, "Active Tunable Absorption Enhancement with Graphene Nanodisk Arrays," Nano letters **14**, 299-304 (2014).
32. W. Luo, W. Cai, Y. Xiang, W. Wu, B. Shi, X. Jiang, N. Zhang, M. Ren, X. Zhang, and J. Xu, "In-Plane Electrical Connectivity and Near-Field Concentration of Isolated Graphene Resonators Realized by Ion Beams," Advanced Materials **29** (2017).
33. A. Narita, X. Feng, Y. Hernandez, S. A. Jensen, M. Bonn, H. Yang, I. A. Verzhbitskiy, C. Casiraghi, M. R. Hansen, A. H. R. Koch, G. Fytas, O. Ivasenko, B. Li, K. S. Mali, T. Balandina, S. Mahesh, S. De Feyter, and K. Muellen, "Synthesis of structurally well-defined and liquid-phase-processable graphene nanoribbons," Nature Chemistry **6**, 126-132 (2014).

34. K. S. Kim, Y. Zhao, H. Jang, S. Y. Lee, J. M. Kim, K. S. Kim, J.-H. Ahn, P. Kim, J.-Y. Choi, and B. H. Hong, Large-Scale Pattern Growth of Graphene Films for Stretchable Transparent Electrodes, Nature 457, 706 (2009).

35. M. König, G. Ruhl, J.-M. Batke, and M. C. Lemme, Self-Organized Growth of Graphene Nanomesh with Increased Gas Sensitivity, Nanoscale 8, 15490 (2016).

36. Z. Fang, S. Thongrattanasiri, A. Schlather, Z. Liu, L. Ma, Y. Wang, P. M. Ajayan, P. Nordlander, N. J. Halas, and F. Javier Garcia de Abajo, "Gated Tunability and Hybridization of Localized Plasmons in Nanostructured Graphene," Acs Nano **7**, 2388-2395 (2013).

37. C. Wang, W. Liu, Z. Li, H. Cheng, Z. Li, S. Chen, and J. Tian, "Dynamically Tunable Deep Subwavelength High-Order Anomalous Reflection Using Graphene Metasurfaces," Advanced Optical Materials **6** (2018).

38. J. Kim, H. Son, D. J. Cho, B. Geng, W. Regan, S. Shi, K. Kim, A. Zettl, Y.-R. Shen, and F. Wang, "Electrical Control of Optical Plasmon Resonance with Graphene," Nano letters **12**, 5598-5602 (2012).

39. N. Kakenov, O. Balci, T. Takan, V. A. Ozkan, H. Altan, and C. Kocabas, *Observation of Gate-Tunable Coherent Perfect Absorption of Terahertz Radiation in Graphene*, ACS Photonics **3**, 1531 (2016).

40. K. S. Novoselov, A. K. Geim, S. V. Morozov, D. Jiang, Y. Zhang, S. V. Dubonos, I. V. Grigorieva, and A. A. Firsov, "Electric field effect in atomically thin carbon films," Science **306**, 666-669 (2004).

41. F. Monticone, C. A. Valagiannopoulos, and A. Alu, "Parity-Time Symmetric Nonlocal Metasurfaces: All-Angle Negative Refraction and Volumetric Imaging," Physical Review X **6** (2016).

42. F. H. L. Koppens, D. E. Chang, and F. Javier Garcia de Abajo, "Graphene Plasmonics: A Platform for Strong Light-Matter Interactions," Nano letters **11**, 3370-3377 (2011).

43. M. Liu, W. Kang, Y. Zhang, H. Zhang, Y. Zhang, and D. Li, "Dynamically controlled terahertz coherent absorber engineered with VO2-integrated Dirac semimetal metamaterials," Optics Communications **503**, 127443 (2022).

44. X. M. Li, H. Feng, M. J. Yun, Z. Wang, Y. G. Hu, Y. J. Gu, F. H. Liu, and W. P. Wu, "Polarization-independent and all-optically modulated multiband metamaterial coherent perfect absorber," Optics and Laser Technology **166** (2023).

45. P. C. Wu, C. Y. Liao, V. Savinov, T. L. Chung, W. T. Chen, Y.-W. Huang, P. R. Wu, Y.-H. Chen, A.-Q. Liu, N. I. Zheludev, and D. P. Tsai, "Optical Anapole Metamaterial," Acs Nano **12**, 1920-1927 (2018).

46. D. J. Cho, F. Wang, X. Zhang, and Y. R. Shen, "Contribution of the electric quadrupole resonance in optical metamaterials," Physical Review B **78** (2008).

47. C. Liu, L. Chen, T. Wu, Y. Liu, R. Ma, J. Li, Z. Yu, H. Ye, and L. Yu, "Characteristics of electric quadrupole and magnetic quadrupole coupling in a symmetric silicon structure," New Journal of Physics **22** (2020).

48. S. V. Morozov, K. S. Novoselov, M. I. Katsnelson, F. Schedin, D. C. Elias, J. A. Jaszczak, and A. K. Geim, "Giant intrinsic carrier mobilities in graphene and its bilayer," Physical review letters **100** (2008).

49. T. Guo, and C. Argyropoulos, "Broadband polarizers based on graphene metasurfaces," Optics Letters **41**, 5592-5595 (2016).

50. D. G. Baranov, A. Krasnok, T. Shegai, A. Alu, and Y. Chong, "Coherent perfect absorbers: linear control of light with light," Nature Reviews Materials **2** (2017).